\documentstyle[11pt,paspconf,epsf]{article}
\newcommand{\eq}{\begin{equation}}
\newcommand{\en}{\end{equation}}
\newcommand{\msun}{M$_\odot$}

\def\teff{$T_eff$}

\begin{document}
\title{Metallicity-dependent Spectral Evolution}
\author{Peeter Traat}
\affil{Tartu Observatory, EE2400 Tartu, Estonia}

\section{Introduction}
Initial chemical composition of stars is, besides the mass, another key  
factor in stellar evolution. Through stellar lifetimes and impact on 
radiation output and nucleosynthesis of stars it is controlling both the pace 
of evolution of galactic matter/light and changes in their integrated 
observables and spectra. 

Given aspect of galactic evolution has been for long 
time ignored in quantitative studies, because the real breaktrough in 
availability of sets of homogeneous non-solar composition stellar evolutionary 
tracks and atmospheres occurred only in the last few years. Presently there 
exist two consistent track sets with sufficient range of metallicities - 
set for 5 compositions produced by Geneva group (cf. Meynet {\it et al.} 
1994, with earlier references therein; lately a sixth, high-metallicity 
$Z=0.10$ 
composition got added to the set by Mowlavi {\it et al.} 1998)
and Padova tracks (
cf. Girardi {\it et al.} 1996, and references therein), 
the latter covering with 8 compositions practically all
the observable and theoretically useful metallicity interval from 
$Z=0.0001$ until $Z=0.10$. Both still pose some further problems for 
applications(ors) - 
Geneva tracks do not include for non-solar compositions past-RGB stages of 
evolution of low-mass stars $M \le 1.7$ \msun, so these have to be added 
from other 
(and inhomogeneous) sources, 
plus it has shorter range in abundances. Padova tracks 
are relatively free of so basic problematics, except of 
lack of stars $M>9$ \msun \, in $Z=0.10$ subset, but some minor 
irregularities arise also for $Z=0.001$ 
composition because they were computed with different radiative opacities. 
As to the coupling of $Z$-dependent chemical evolution with spectrophotometric, 
Geneva tracks have had the preference of having stellar nucleosynthetic 
results available from the onset (Maeder 1992), although 
for just two compositions, $Z=0.001$ and $Z=0.02$. That gap 
of missing yields for Padova set got finally filled by a recent 
publication of Portinari {\it et al.} (1998), for first time putting the future 
combined chemospectrophotometric models on firm leg.
\begin{table}
\vspace*{-.4cm}
{\bf Table 1. Time-integrated energy fluxes} of coeval stellar populations
with different IMF slopes and metallicities, per 1 M$_\odot$ in massive 
stars $10 \div 120$  M$_\odot$.  
Flux units  $10^{52}$ erg, flux fractions f$_{\rm i}$ computed for 
intervals 90.9-395 \AA, 395-995 \AA, 995-1995 \AA, 1995-3590 \AA, 
3590-7490 \AA, 0.749-4.99 $\mu$m and 4.99-160 $\mu$m. \hfill
\vspace{.8ex}
\centering{\small
\begin{tabular}{c|c|cccccccc} 
 \hline 
\mbox{\rule[-.9ex]{0ex}{3.2ex} n} & Z & L & f$_1$ & f$_2$ & f$_3$ & f$_4$ & f$_5$ & f$_6$ & f$_7$ \\ 
\hline 
\mbox{\rule{0ex}{2.2ex} 1.0}  & 0.001 &  0.2944 &.0792 &.2585 &.4691 &.1090 &.0635 &.0206 &.0002\\
     & 0.004 &  0.2977 &.0503 &.2429 &.4932 &.1153 &.0695 &.0284 &.0002\\
     & 0.008 &  0.2896 &.0383 &.2465 &.5128 &.1170 &.0590 &.0262 &.0003\\
     & 0.020 &  0.2629 &.0243 &.2408 &.5265 &.1172 &.0539 &.0368 &.0005\\
     & 0.040 &  0.2305 &.0096 &.2012 &.5476 &.1367 &.0659 &.0385 &.0005\\
\hline 
1.6  & 0.001 &  0.3169 & .0579 & .2025 & .4809 & .1239 & .0911 & .0434 & .0004\\
     & 0.004 &  0.3147 & .0365 & .1902 & .4896 & .1263 & .0988 & .0579 & .0006\\
     & 0.008 &  0.3074 & .0271 & .1880 & .5032 & .1287 & .0919 & .0603 & .0007\\
     & 0.020 &  0.2824 & .0166 & .1799 & .5020 & .1256 & .0888 & .0859 & .0012\\
     & 0.040 &  0.2487 & .0067 & .1527 & .5175 & .1412 & .0962 & .0844 & .0012\\
\hline 
2.0  & 0.001 &  0.3897 & .0376 & .1405 & .4561 & .1388 & .1393 & .0868 & .0009\\
     & 0.004 &  0.3746 & .0239 & .1339 & .4546 & .1365 & .1444 & .1055 & .0012\\
     & 0.008 &  0.3678 & .0174 & .1296 & .4578 & .1375 & .1420 & .1144 & .0014\\
     & 0.020 &  0.3458 & .0103 & .1201 & .4352 & .1291 & .1419 & .1611 & .0023\\
     & 0.040 &  0.2997 & .0043 & .1052 & .4519 & .1415 & .1441 & .1509 & .0021\\
\hline 
2.2  & 0.001 &  0.4691 & .0274 & .1064 & .4253 & .1459 & .1741 & .1197 & .0013\\
     & 0.004 &  0.4399 & .0176 & .1027 & .4200 & .1411 & .1772 & .1397 & .0016\\
     & 0.008 &  0.4337 & .0127 & .0982 & .4162 & .1403 & .1775 & .1533 & .0019\\
     & 0.020 &  0.4159 & .0073 & .0888 & .3811 & .1278 & .1778 & .2141 & .0030\\
     & 0.040 &  0.3538 & .0032 & .0799 & .4000 & .1388 & .1782 & .1973 & .0027\\
\hline 
2.35 & 0.001 &  0.5620 & .0205 & .0823 & .3947 & .1502 & .2029 & .1477 & .0016\\
     & 0.004 &  0.5159 & .0133 & .0805 & .3872 & .1436 & .2049 & .1684 & .0019\\
     & 0.008 &  0.5106 & .0095 & .0762 & .3779 & .1411 & .2070 & .1861 & .0023\\
     & 0.020 &  0.4985 & .0053 & .0674 & .3346 & .1249 & .2063 & .2578 & .0037\\
     & 0.040 &  0.4166 & .0024 & .0621 & .3547 & .1350 & .2066 & .2359 & .0032\\
\hline 
2.5  & 0.001 &  0.6974 & .0148 & .0613 & .3599 & .1532 & .2319 & .1769 & .0020\\
     & 0.004 &  0.6256 & .0097 & .0607 & .3505 & .1452 & .2336 & .1980 & .0023\\
     & 0.008 &  0.6221 & .0068 & .0568 & .3360 & .1405 & .2372 & .2201 & .0027\\
     & 0.020 &  0.6198 & .0037 & .0490 & .2864 & .1204 & .2340 & .3021 & .0043\\
     & 0.040 &  0.5075 & .0017 & .0464 & .3071 & .1298 & .2355 & .2757 & .0037\\
\hline 
2.7  & 0.001 &  0.9767 & .0090 & .0391 & .3115 & .1551 & .2681 & .2148 & .0024\\
     & 0.004 &  0.8492 & .0060 & .0394 & .2997 & .1458 & .2706 & .2358 & .0027\\
     & 0.008 &  0.8506 & .0041 & .0363 & .2793 & .1377 & .2754 & .2640 & .0032\\
     & 0.020 &  0.8719 & .0022 & .0303 & .2255 & .1126 & .2666 & .3577 & .0050\\
     & 0.040 &  0.6937 & .0010 & .0297 & .2451 & .1211 & .2717 & .3270 & .0044\\
\hline 
3.0  & 0.001 &  1.7551 & .0038 & .0183 & .2447 & .1540 & .3122 & .2641 & .0030\\
     & 0.004 &  1.4602 & .0026 & .0188 & .2295 & .1437 & .3183 & .2839 & .0032\\
     & 0.008 &  1.4817 & .0017 & .0169 & .2035 & .1306 & .3229 & .3206 & .0038\\
     & 0.020 &  1.5820 & .0009 & .0133 & .1510 & .0993 & .3025 & .4269 & .0060\\
     & 0.040 &  1.2094 & .0004 & .0138 & .1660 & .1063 & .3152 & .3931 & .0052\\
\hline 
3.4  & 0.001 &  4.2012 & .0011 & .0060 & .1768 & .1485 & .3510 & .3130 & .0036\\
     & 0.004 &  3.3257 & .0007 & .0063 & .1585 & .1382 & .3636 & .3290 & .0037\\
     & 0.008 &  3.4400 & .0005 & .0054 & .1303 & .1191 & .3651 & .3752 & .0044\\
     & 0.020 &  3.8395 & .0002 & .0040 & .0867 & .0828 & .3286 & .4907 & .0069\\
     & 0.040 &  2.8204 & .0001 & .0044 & .0948 & .0875 & .3505 & .4567 & .0060\\
\hline 
4.0  & 0.001 & 16.8714 & .0002 & .0010 & .1147 & .1383 & .3811 & .3604 & .0043\\
     & 0.004 & 12.6020 & .0001 & .0011 & .0950 & .1293 & .4026 & .3678 & .0041\\
     & 0.008 & 13.4144 & .0001 & .0009 & .0693 & .1037 & .3968 & .4243 & .0050\\
     & 0.020 & 15.6599 & .0000 & .0006 & .0396 & .0647 & .3416 & .5457 & .0078\\
     & 0.040 & 11.1299 & .0000 & .0007 & .0415 & .0662 & .3711 & .5137 & .0067\\
\hline
\end{tabular}}
\end{table} 
\section{Discussion of spectral data}
We have used both track sets to study spectral evolution of stellar 
populations in function of their initial metallicity and star formation 
parameters. No gas emission/dust absorption has been considered 
 in these models, since it is highly individual for galaxies. 
 On the basis of Padova tracks and isochrones 
     we have some time ago 
     released a standardized grid of spectral data files (Traat 
     1996), computed with the evolutionary codes developed by the 
     author.
  The grid includes 1008 models and has the widest practically useful 
 range of chemical compositions and star formation prescriptions. It  
 covers all the Padova set abundance range, i.e. eighth metallicities 
 $Z = 0.0001$, 0.0004, 0.001, 0.004, 0.008, 0.02, 0.05 and 0.10.  
  The conversion from stellar luminosities and temperatures to spectral flux 
  distributions was  
  based on Kurucz (1993) model atmospheres with the prospect of their 
  future replacement in cool star region \teff$ \le 4500 {}^\circ\!$K to 
  new Uppsala release. 
  
Dataset for each $Z$ value includes
 \begin{itemize}
\vspace{-1.5ex}
    \item six power law IMF-s of different slopes $1.6 \div 3.5$, 
\vspace{-1.5ex}
    \item corresponding single-generation populations
\vspace{-1.5ex}
    \item
    populations with continuous/continuing star formation for those IMF-s 
    with 4 SFR index 
    values $s=0$ (constant SFR), $s=1$ (exponentially declining), $s=1.5$ and 
    2 (initially faster, later slower than the exponential SFR) and 5 
    star-formation timescales 0.2, 1, 2, 5 and 15 Gy. 
 \end{itemize}
\vspace{-1.5ex}
    Spectra are presented for 50 ages in the case of starburst populations 
and 20 ages for populations with continuous star formation. Ages range from 
0.003 Gy to 20 Gy. 

     Models are chemically homogeneous, no 
  nebular component/absorption has been included. SFR has been parametrized 
  by a power of the gas volume density (as introduced by Schmidt (1959)), with 
  index $s$ and time-scale $t_0$. In this context, the single-generation 
  ("initial-burst") populations with their independency on the SFR/its power 
  index form the limiting "starburst" $t_0 = 0$ case. 

Metallicity growth affects the stellar evolution in two ways: first, making 
stars cooler and dimmer, secondly, redistributing their flux through opacity growth 
at short wavelengths towards longer wavelengths. 
So the summatic result of $Z$ rise on the composite spectrum of a stellar population 
is the progressive erosion of flux in ultraviolet region, and the faster, the 
shorter the wavelength. However, with and the enhancement of
absorption in metallic lines in UV the bulk of energy reradiated in the 
optical and near-infrared is increasing with the definite net result that 
the spectrum level at $\lambda \geq 1$ $\mu$m is for all population ages 
progressively higher for higher metallicities. 

Table 1 gives a general quantitative review of the extent of metallicity 
effects in different spectral regions for populations with different mass
function slopes $n$. Time is eliminated by integration over the lifetimes of 
stars,  population mass is scaled to the unit amount of mass in very massive 
stars $M\ge 10$ \msun, Geneva set of tracks (5 compositions, the latest 
$Z=0.10$ subset was not included yet) was used. These data testify, that 
composition-caused flux changes can be rather impressive, extending to factor 
of 10 in far-UV and $\sim 2$ in infrared. 

As to illustrate metallicity effects graphically, we also provide a couple of 
examples on Fig. 1, computed with Padova tracks. On the left panel of this figure 
we have plotted computed spectra 
of coevally formed stellar populations (so-called {\it initial burst}  populations) 
with 8 different initial metallicities $Z$, having ages 1 Gy. Such comparatively blue spectra 
are typical for medium-aged 
star clusters of assumed chemical compositions,    
or young elliptical galaxies with ages somewhat exceeding 1 Gy. 
The flux of models is scaled 
to the unit mass in luminous stars with masses $M > 0.6$ \msun, 
the IMF in these graphs is a power-law with slope $n = -2.35$ 
("Salpeter" value). The most metal-deficient, $Z=0.0001$ spectrum has 
a correct location in log $E_\lambda$, all the others have been 
successively shifted downwards by 
additional 0.5 dex, with maximum total shift $-3.5$ 
for the $Z=0.10$ case. 
On the right panel
the spectra of old, 10 Gy stellar populations, are plotted, in which star formation 
is continuous and proceeds
 with a constant, time-independent intensity over the 
eon $t_0$. Given case might be considered as a kind of approximation to late spirals 
or irregulars,
since in many of these actual SFR-s do not seem to 
significantly differ from the mean average over their past. 
Notice, however, that due to 
the actively continuing star formation the ultraviolet flux of models keeps 
 sizable. The growth of metal content sharply reduces the flux at shorter 
wavelengths, as also on the left-panel plot. 
\begin{figure}
\plottwo{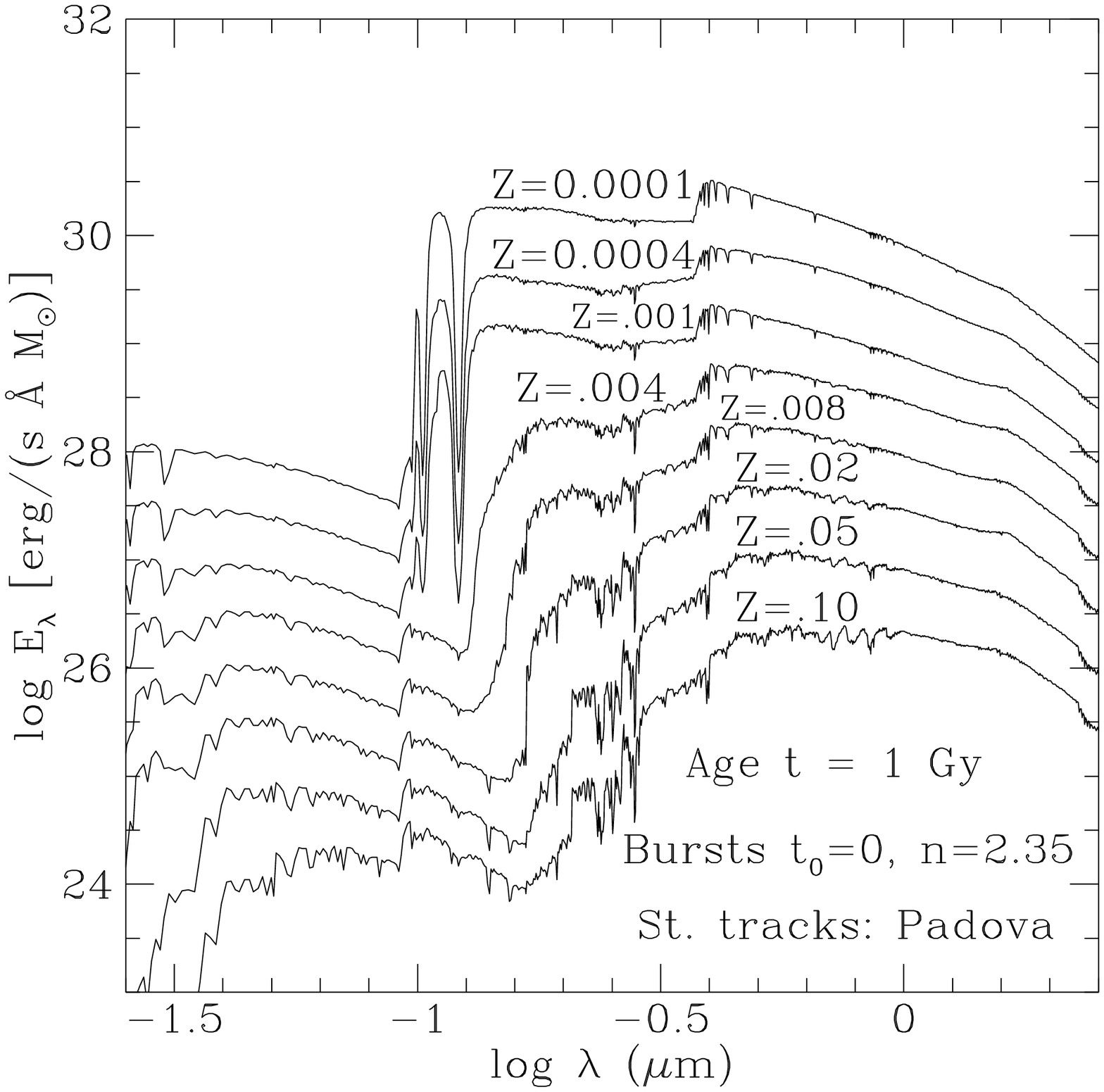}{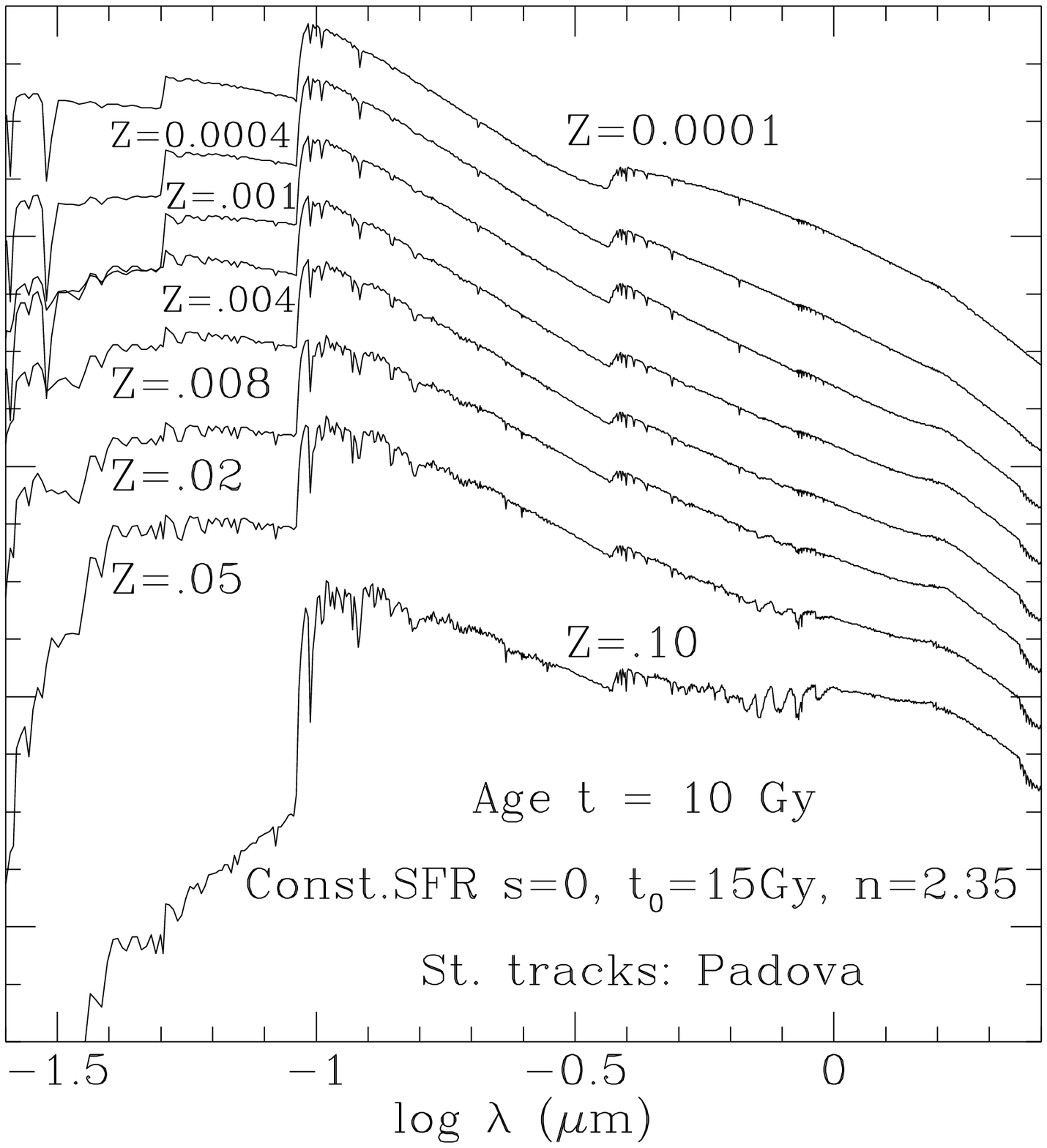}
\caption{Composition dependence of spectra of stellar populations: coevally formed 
young stellar populations (age 1 Gy, {\it left panel}) and old populations with constant SFR
(age 10 Gy, {\it right panel}).}
\end{figure}

\acknowledgments
I would like to express my gratitude to the LOC of Symposium and Tartu Cultural 
Capital for supports facilitating my participation in the meeting.

\end{document}